\documentclass[18pt]{article}
\begin{document}
\title{Simultaneous emergence of curved  spacetime and quantum mechanics}

 \author{\textbf{S S De$^1$ and F Rahaman$^2$} \\
\\
 $^1$Department of Applied Mathematics,  University of Calcutta,
Kolkata -  700009, India\\
E-mail:  ssddadai08@rediffmail.com \\
\\
$^2$Department of   Mathematics,  Jadavpur University,
Kolkata -  700032, India\\
E-mail:  rahaman@iucaa.ernet.in}

\maketitle

\begin{abstract}

It is shown in this paper that the geometrically structure-less
spacetime manifold is converted instantaneously to a curved one,
the Riemannian or may be  a Finslerian spacetime with an
associated Riemannian spacetime, on the appearance of quantum
Weyl spinors dependent only on time in that background flat
manifold and having the symplectic  property in the abstract space
of spinors.  The scenario depicts simultaneous emergence of the
gravity in accord with general relativity and quantum
mechanics.The emergent gravity leads to the generalized
uncertainty principle, which in turn, ushers in discrete space
time. The emerged space time is specified here as to be
Finslerian  and the field equation in that space time has been
obtained from the classical one due to the arising quantized
space and time. From this field equation we find the quantum
field equation for highly massive (of the Planck order) spinors in
the associated Riemannian space of the Finsler space, which is in
fact, the background homogeneous and isotropic FRW space time of
the universe.These highly massive spinors provide the mass
distribution complying Einstein equivalence principle. All these
occurred in the indivisible minimum time considered as zero time
or spontaneity.

\end{abstract}

Pacs No.:  04.60.-m , 98.80.-k, 98.80.Jk
\\
\\

\section {Introduction}

 \par One of the crucial problems in contemporary theoretical physics is to unify the four
  known interaction forces.
 After the advent of the Grand Unified Theories (GUT) the problem is now shifted in
 establishing a
  successful relation between the relativistic quantum field theories formulated on
   Minkowski space
  time,
  which are providing the conceptual framework of standard model of particle physics
  and the general covariance
   of space time in gravity according to the general relativity (G R). In fact, to achieve
    a veritable unification of
   G R and Quantum mechanics (Q M) it may not be necessary to have any presupposition of
   geometrical structure in the
    space time manifold or even smoothness of it. Also the physics of the Planck regime
    of the universe should be understood
    for this purpose. Thus the  problem    is that  how the gravity might have been emerged as the
     metric structure of space time manifold
    in the usual setting of G R (being the origin of the Einstein equivalence principle).

For establishing the unification and also accomplishing
quantization of gravity, there are several approaches such as
string theory, loop quantum theory, theories based on discreteness
of space time  [ 1-9]. In these
theories either fundamental status of Q M or that of gravity has
been postulated.  Consequently, we have to encounter another
     important problem closely related to the unification of G R and Q
     M.
This is being to ascertain the fundamental status to any one of G R
and Q M or to both of them.In case one of them is fundamental, the
other must be emergent.   But a perfect unification should demand
that both of them must be fundamental as well as emergent.
Consequently, we have to encounter another
     important problem closely related to the unification of G R and Q
     M.
This is being to ascertain the fundamental status to any one of G
R and Q M,or to both of them.In case one of them is fundamental,
the other must be emergent. But a perfect unification should
demand that both of them must be fundamental as well as emergent.
Also, we have to specify the actual phenomenon of their emergence,
 preferably, simultaneous emergence.
To this end, it   is presently proposed here that neither of Q M and
gravity can be taken as to be fundamental, and on the contrary,
their simultaneous and spontaneous emergence has been considered in
a geometrically structure-less background space time manifold with
the appearance of quantum Weyl spinors dependent only on time and
having a symplectic  property in the abstract space of spinors. In
fact, apparition of these spinors can instantaneously convert the
space time manifold into a metric structured one as well as to a
discrete- structured space time manifold having minimum space and
time lengths, all of the Planck scale. Thus, gravity in the setting
of G R emerges together with Q M manifested by the quanta of space
and time and also by the quantum spinors. The consequence of the
minimum time and space lengths is the generalized uncertainty
principle (G U P) and vice versa. That is, the Heisenberg
uncertainty principle is to be modified to G U P due to the emergent
gravity and this must lead to the minimum time and space lengths. In
this indivisible minimum time both Q M and G R emerged and after
this quantum of time considered as "instantaneous", the time
discreteness has no bearing on the subsequent evolution of the
universe. Highly massive particles might have been generated in this
Planck order minimum time and made it possible to comply Einstein
equivalence principle.

   We begin in section II, with the discussion on the emergence of structured space time
    due to the appearance of Weyl spinors in that flat space time
   manifold.
    In section III, identification of the emerged space time has been made. In section
    IV, in an alternative approach,
    the metric tensor of this space time is
     constructed from the basis spinors.  In section V, a brief discussion has been made on
      GUP as originated
from gravitational interaction of photon with electron.  We
discuss in section VI, the appearance of
      highly massive particles as obtained from the field
     equation in associated Riemannian space, which is derived from the field equation
      in the emerged Finsler space. In final section
      VII the full scenario is presented with some concluding remarks.

\section{ EMERGENCE OF STRUCTURED SPACE TIME :}

It is assumed here the appearance of two independent two-dimensional spinor fields
 depending only on time in the flat background space
time manifold. The space of the two dimensional spinors is supposed to be endowed
with a symplectic  structure induced by a skew symmetric scalar product.
 This spinors $\psi^{(1)}$ and $\psi^{(2)}$ are in fact Weyl spinors, that is, they
  satisfy Weyl equation. If these spinors would have been dependent
  on both the space and time coordinates of the manifold, then the manifold may remain
   geometrically structure-less or flat. On the other hand,
   if they depend only on time, then the space time manifold must spontaneously have a
curved  structure, and consequently a non zero correction to the
   spinor fields arises. With this non zero correction $\Gamma_{\mu}$ or  the generalized derivative (covariant derivative) $\textit{D}_{\mu}$ ,
    we have the Weyl equation as the desired equation for the
   spinors in the curved space time. Here, the symplectic structure in the two dimensional
    spinor space, as induced by the skew-symmetric scalar product
   denoted by [.,.] is to be understood as the relation \\
\begin{equation}[\phi ,\chi]=-[\chi ,\phi],\end{equation}
must remain valid for the spinors $\phi$, $\chi$. This scalar product can be expressible
 by a matrix\\
\begin{equation}\epsilon_{\alpha \beta}=\left(
                            \begin{array}{cc}
                              0 & 1 \\
                              -1 & 0 \\
                            \end{array}
                          \right)\end{equation}
and with the components of the spinors (referred to a certain basis). The spinors
are expressed as \\
\begin{equation}\phi=\left(
         \begin{array}{c}
           \phi^{1} \\
           \phi^{2} \\
         \end{array}
       \right),
 \chi=\left(
        \begin{array}{c}
          \chi^{1} \\
          \chi^{2} \\
        \end{array}
      \right). \end{equation}
The scalar product is \\
\begin{equation}[\phi ,\chi]=\epsilon_{\alpha \beta} \phi^{\alpha} \chi^{\beta}=\left(
                                                                     \begin{array}{cc}
                                                                       \phi^{1} & \chi^{1} \\
                                                                       \phi^{2} & \chi^{2} \\
                                                                     \end{array}
                                                                   \right). \end{equation}
The matrix (2) raises or lowers the indices of the spinors as follows:\\
\begin{equation}\phi^{\alpha}=\epsilon^{\alpha \beta} \phi_{\beta},\phi_{\alpha}=
\epsilon_{\alpha \beta} \phi^{\beta},\end{equation}
where
\begin{equation}\epsilon^{ \alpha \beta}=\left(
                             \begin{array}{cc}
                               0 & -1 \\
                               1 & 0 \\
                             \end{array}
                           \right)=\epsilon^{T}_{\alpha \beta}=\epsilon^{-1}_{\alpha
                           \beta}.\end{equation}
Also, we have
\begin{equation} [\phi,\chi]=-\epsilon^{\alpha \beta}\phi_{\alpha} \chi_{\beta}=
-[\chi,\phi].\end{equation}
\\
As it is said earlier, the two independent spinors $\psi
^{(1)},\psi ^{(2)}$ are Weyl spinors and satisfy the Weyl
equation. This Weyl equation can be obtained from the following
Dirac equation in the curved space time for a massless particle
or field [ 10-12 ] as
\\
\begin{equation}i\gamma ^{a}e^{\mu}_{a}\left( \partial
_{\mu}+\hat{\Gamma_{\mu}}\right)\hat{\psi^{(i)}}=0, \end{equation}
\\
where $e^{\mu}_{a}$ denotes the vierbein and $\gamma^{\mu}$
 are related to the flat space $\gamma^{a}$ matrices by the relation
$\gamma^{\mu}=e^{\mu}_{a}\gamma^{a}$.
  The spin connection $\hat{\Gamma_{\mu}}$ is given by \\
 \begin{equation}\hat{\Gamma_{\mu}}=\frac{1}{2}e^{\gamma}_{h}\nabla_{\mu}e_{\gamma
k}S^{hk}=\frac{1}{2}e^{\gamma}_{h}\left( \partial_{\mu}e_{\gamma
k}-\Gamma^{\sigma}_{\mu \nu}e_{\sigma
k}\right)S^{hk}. \end{equation}
\\
Here, $S^{hk}=\frac{1}{4}\left[\gamma^{h},\gamma^{k}\right]$ is the skew symmetric
generator of the Lorentz transformation for the spinor field and $\Gamma^{\sigma}_{\mu \nu}$
 is the affine connection. By choosing the Dirac-Pauli representation of $\gamma^{a}$
 matrices as\\
\[ \gamma^{0}=\left(
              \begin{array}{cc}
                1 & 0 \\
                0 & -1 \\
              \end{array}
            \right) ,~
 \gamma^{i}=\left(
              \begin{array}{cc}
                0 & -\sigma^{i} \\
                \sigma^{i} & 0 \\
              \end{array}
            \right) ; i=1,2,3,  \]
where $\sigma^{i}$ are Pauli matrices, we can find after a straight forward
 calculation the following Weyl equations for the Weyl spinor field $\psi^{(i)}$
 by writing the Dirac spinor $\hat{\psi^{(i)}}$ as
$\hat{\psi^{(i)}}=\left(
                    \begin{array}{c}
                      \psi^{(i)} \\
                      -\psi^{(i)} \\
                    \end{array}
                  \right).$\\
It is given as\\
\begin{equation}\sigma^{a}e^{\mu}_{a}\left(\partial_{\mu}+\Gamma_{\mu}\right)\psi^{(i)}
=0,\end{equation}
where $\sigma^{0}=1$ and \\
\begin{equation}\Gamma_{\mu}=-\frac{1}{4}\left[\sum_{h,k=1,2,3}e^{\beta}_{h}
\nabla_{\mu}e_{\beta k}
\sigma^{h}\sigma^{k}+\sum_{k=1,2,3}\sigma^{k}\left(e^{\beta}_{k}
\nabla_{\mu}e_{\beta 0}-e^{\beta}_{0}\nabla_{\mu}e_{\beta
k}\right)\right].\end{equation}
The Weyl equation can also be written as\\
\begin{equation}i\sigma^{\mu}\left(\partial_{\mu}+\Gamma_{\mu}\right)\psi^{(i)}=0,\end{equation}
where $\sigma^{\mu}=e^{\mu}_{a}\sigma^{a}$.\\

By choosing the representation of $\gamma^{a}$ matrices as
\[ \gamma^{0}=\left(
              \begin{array}{cc}
                1 & 0 \\
                0 & -1 \\
              \end{array}
            \right) ,~~
 \gamma^{i}=\left(
              \begin{array}{cc}
                0 & \sigma^{i} \\
                -\sigma^{i} & 0 \\
              \end{array}
            \right), i=1,2,3 ,\]
we can have the Weyl equation as\\
\begin{equation}i\left(
\sigma^{0}e^{\mu}_{0}-\sum_{k=1,2,3}\sigma^{k}e^{\mu}_{k}\right)
\left(\partial_{\mu}+\Gamma_{\mu}\right)\psi^{(i)}=0, \end{equation}
where
\begin{equation}\Gamma_{\mu}=-\frac{1}{4}\left[\sum_{h,k=1,2,3}e^{\beta}_{h}
\nabla_{\mu}e_{\beta
k}\sigma^{h}\sigma^{k}-\sum_{k=1,2,3}\sigma^{k} \left(
e^{\beta}_{k}\nabla_{\mu}e_{\beta
0}-e^{\beta}_{0}\nabla_{\mu}e_{\beta k}\right)
\right].\end{equation}

For the case of $\Gamma_{a}=0$,   $\psi^{(i)}$ must  depend on
both the
 space and time co ordinates of the manifold and vice versa, that is, space and time
  dependent spinors can satisfy Weyl equation with vanishing $\Gamma_{a}$
  term. In this case, the flat space time manifold remains flat. On the contrary,
   if $\psi^{(i)}$ depends only on time (but not a constant), then it must have
  to satisfy Weyl equation with the non zero $\Gamma_{a}$, which can also be written
( follows from  both the equations (12) and (14) )
\begin{equation}i \sigma^{0}\partial_{0}\psi^{(i)}_{\alpha}+i \sigma^{a}\Gamma_{a \alpha \beta}
\psi^{(i)}_{\beta}=0, i=1,2  ,  \end{equation}
since $\psi^{(i)}_{\alpha}$ depends only on time $x^{0}=c t$. The non zero $\Gamma_{a}$
 implies that the space time manifold must have changed to a curved one.

\par In this connection it is to be mentioned here that Kober (2008) has constructed
 "connection" of the spinor fields by defining a covariant derivative
in such a way that it leaves two independent space time dependent
basis spinors constant. With this covariant derivative the basis
spinor fields automatically satisfy Weyl equation in curved space
time. Also a tetrad formulation and metric with Minkowski
signature have been constructed from the spinors having symplectic
structure in the abstract space of spinors. But the resulted
curved space time manifold may not have the same covariant
derivative (as define therein ), that is, corresponding curved
space time of the covariant derivative may not be the curved
space time obtained from tetrad formulation. In fact, it was there
only a
 presupposition of metric structure for the space time manifold in
  which the "connection" or covariant derivative are all defined.\\
\par By using symplectic  condition (1) for the spinors $\psi^{(1)}$
 and $\psi^{(2)}$,  we can have from the equation (15) as
\begin{equation}\sigma^{a}\Gamma_{a \alpha \beta}=\frac{\partial_{0}
\psi^{(i)}_{\alpha}\psi^{(2)\beta}-\partial_{0}\psi^{(2)}_{\alpha}\psi^{(1)\beta}}
{[\psi^{(1)},\psi^{(2)}]} ,\end{equation}
where $\psi^{(1)}_{\alpha},\psi^{(1)\beta}$ etc  are related by
 $\epsilon_{\alpha \beta}$ as in (5). Here, $\sigma^{a}\Gamma_{a \alpha \beta}$ is
  dependent
on the spinors $\psi^{(1)}$ and $\psi^{(2)}$ and has the similar form for the
 spin connection defined by [13].\\ \\

\section{ SPECIFICATION OF THE EMERGED SPACE TIME :}

\par The independent spinors $\psi^{(1)}$ and $\psi^{(2)}$ are specifically
 taken as $q(x^{0})\left(
   \begin{array}{c}
                    1 \\
                              0 \\
                                   \end{array}
                                      \right)
$ and $q(x^{0})\left(
                 \begin{array}{c}
                   0 \\
                   1 \\
                 \end{array}
               \right)
$ (referred to certain basis) respectively. These are dependent only on time $x^0$ and we have\\
\begin{equation}\psi^{(1)}=\left(
              \begin{array}{c}
                \psi^{(1)}_{1} \\
                \psi^{(1)}_{2} \\
              \end{array}
            \right)
=\left(
    \begin{array}{c}
      q(x^0) \\
      0 \\
    \end{array}
  \right)
  ~and~
\psi^{(2)}=\left(
             \begin{array}{c}
               \psi^{(2)}_{1} \\
               \psi^{(2)}_{2} \\
             \end{array}
           \right)
 =\left(
    \begin{array}{c}
      0 \\
      q(x^0) \\
    \end{array}
  \right). \end{equation}
  These Weyl spinors are also the spinors in the sense of Cartan [14].
  A spinor $(\xi_{0},\xi_{1})$ is a
   spinor according to Cartan if \[ x^{2}_{1}+x^{2}_{2}+x^{2}_{3}=0,\]  where
    \[ x_{1}=\xi^{2}_{0}-\xi^{2}_{1}, ~x_{2}=i(\xi^{2}_{0}+\xi^{2}_{1})\]
    and \[ x_{3}=-2\xi_{0}\xi_{1}, \]
for \[ \psi^{(1)}~=~\xi_{0} = q(x^0)~~ and ~~ \xi_{1}=0.\]  Consequently,
\[ x_{1}=q(x^0)^2, x_{2}=iq(x^0)^2, x_{3}=0.\]
  Therefore, \[ x^{2}_{1}+x^{2}_{2}+x^{2}_{3}=0.\]  Similarly, we see that
   $\psi^{(2)}$ is a spinor according to Cartan. In fact, any two component
  Weyl spinor is a Cartan spinor too.\\ \\
  \par Now, using (5), we have\\
\begin{equation}\left(
      \begin{array}{c}
        \psi^{(1)1} \\
        \psi^{(1)2} \\
      \end{array}
    \right)=\left(
              \begin{array}{c}
                0 \\
                q(x^0) \\
              \end{array}
            \right)
            ~~and~~
           \left(
                \begin{array}{c}
                  \psi^{(2)1} \\
                  \psi^{(2)2} \\
                \end{array}
              \right)=\left(
                        \begin{array}{c}
                          -q(x^0) \\
                          0 \\
                        \end{array}
                      \right).\end{equation}\\
Then we have from (16) \\
\begin{equation}\sigma^{a} \Gamma_{a11}=-\frac{q^{'}(x^0)}{q(x^0)}, ~\sigma^{a}
 \Gamma_{a22}=-\frac{q^{'}(x^0)}{q(x^0)}, ~\sigma^{a} \Gamma_{a12}=\sigma^{a}
\Gamma_{a21}=0 .\end{equation}\\
Therefore,\\
\begin{equation}\left(
    \begin{array}{c}
      \sigma^{a} \Gamma_{a \alpha \beta} \\
    \end{array}
  \right)=-\frac{q^{'}(x^0)}{q(x^0)} \left(
                                       \begin{array}{cc}
                                         1 & 0 \\
                                         0 & 1 \\
                                       \end{array}
                                     \right)=-\frac{q^{'}(x^0)}{q(x^0)}I .\end{equation}
Then  the Weyl equation (15) becomes\\
\[\sigma^{0}\partial_{0}\psi^{(i)}_{\alpha}-\frac{q^{'}(x^0)}{q(x^0)}\delta_{\alpha \beta}
\psi^{(i)}_{\beta}=0,\]\\
or\\
\begin{equation}\sigma^{0}\partial_{0}\psi^{(i)}_{\alpha}-\frac{q^{'}(x^0)}{q(x^0)}
\psi^{(i)}_{\alpha}=0,  i=1,2. \end{equation}\\

This equation represents the Weyl equation in the emerged curved
space time due to the appearance of two independent spinors
depending only
 on time. But the exact nature of the metric structure of the manifold is yet
  to be determined.This can be accomplished by a comparison with the
  Weyl equation in a specific Riemannian space time
  manifold. Here we find the Weyl equation for the homogeneous and
  isotropic FRW spacetime ( with flat space ).
The  metric tensor for this spacetime is given by
\begin{equation}g_{00}=1, ~g_{ij}=-a^{2}\delta_{ij}.\end{equation}
The vierbien and inverse vierbein fields are given by \\
\[e^{\mu}_{0}=\delta^{\mu}_{0}, ~e^{\mu}_{i}=\frac{1}{a}\delta^{\mu}_{i},
e^{0}_{\mu}=\delta^{0}_{\mu}, ~e^{i}_{\mu}=a\delta^{i}_{\mu},\]
where $a(x^{0})$ is the scale factor, $x^{0}=ct$ and $\delta^{i}_{\mu}=\delta_{i\mu}.$\\

The affine connection \[ \Gamma^{\sigma}_{\mu \beta}=\frac{1}{2}g^{\sigma \rho}\left
( \partial_{\mu}g_{\beta \rho}+\partial_{\beta}g_{\rho \mu}-\partial_{\rho}g_
{\mu \beta}\right)\] for this space time are given as (in the unit $c=\hbar=1$)\\
\begin{equation}\Gamma^{0}_{0\mu}=0,~~\Gamma^{0}_{ij}=a\dot{a}\delta_{ij},~~
\Gamma^{i}_{0j}=\Gamma^{i}_{j0}=\frac{\dot{a}}{a}\delta_{ij},~~\Gamma^{i}_{jl}=0.
\end{equation}
Consequently, we can find
\[ \sigma^{a}e^{\mu}_{a}\Gamma_{\mu}=\frac{3\dot{a}}{2a},\]  and also
\[ \sigma^{0}
e^{\mu}_{0}-\sum_{k=1,2,3}\sigma^{k}e^{\mu}_{k}=\frac{3\dot{a}}{2a}.\]

Thus we have from (10)and (13), the Weyl equations as
\begin{equation}\sigma^{a}e^{\mu}_{a}\partial_{\mu}\psi^{(i)}+\frac{3\dot{a}}{2a}\psi^{(i)}=0,
\end{equation}
and\\
\[~~~~~~~~~~~~~~~~~~~~~~~~~~~~~~\left(\sigma^{0}e^{\mu}_{0}-\sum_{k=1,2,3}
\sigma^{k}e^{\mu}_{k}\right)
\partial_{\mu}\psi^{(i)}+
\frac{3\dot{a}}{2a}\psi^{(i)}=0~~~~~~~~~~~~~~~~~~~~~~~
~~~~~~~~~~~~~~~~(24 a).\] For $\psi^{(i)}$ dependent only on time
we have the same Weyl
 equation for $\psi^{(i)}$ from (24)and (24 a)\\
\begin{equation}\sigma^{0}\partial_{0}\psi^{i}+\frac{3\dot{a}}{2a}\psi^{(i)}=0.\end{equation}

This equation can be identified with the equation (21) in which we
take
\begin{equation}q(x^0)=\left[a(x^0)\right]^{-\frac{3}{2}},  \end{equation} that
is, ( in unit $c=\hbar =1$ )
\begin{equation}\frac{q^{'}(x^0)}{q^(x^0)}=-\frac{3}{2}\frac{\dot{a}}{a} \end{equation}\\
Therefore, the emerged Riemannian space time manifold due to the
apparition of Weyl spinors which depend only on time and having
symplectic property in the abstract spinor space must be the FRW
space time (the cosmological space time).  It is to be noted that
this FRW spacetime is equivalent to the metric tensor
\begin{equation}g_{\mu \nu}= g(x^0)\eta_{\mu \nu} \end{equation}
In fact, we can achieve this by a pure time transformation
\begin{equation}T= \int \frac{dt}{a(t)} ~~with~~a(t)~\equiv ~\sqrt{g(cT)} = \sqrt{g(x^0)}\end{equation}
where $T$ is the cosmological time. Thus, the FRW space time
(with flat space), the cosmological background space time of the
universe,emerged.In this way,the gravity emerged
spontaneously with Q M which is manifested in the quantum Weyl fields.\\ \\
\section{METRIC STRUCTURE: FINSLER SPACE}

\par We now present here an alternative  construction of space time manifold with
 the metric structure from the two independent time dependent two
component spinors.This consideration is similar to that of [13]  and others
 [15-16]  but it is also different
 in some respects.The two independent spinors are given in (11), from which the four
  vectors in Minkowski space time can be constructed. These are\\
\[ a^{m}=\psi^{(1)\dagger}\sigma^{m}\psi^{(1)}=[q(x^0)]^{2}\sigma^{m}_{11}\]\\
\[ b^{m}=\psi^{(2)\dagger}\sigma^{m}\psi^{(2)}=[q(x^0)]^{2}\sigma^{m}_{22}\]\\
\[ c^{m}=\psi^{(1)\dagger}\sigma^{m}\psi^{(2)}=[q(x^0)]^{2}\sigma^{m}_{12}\]\\
\begin{equation} d^{m}=\psi^{(2)\dagger}\sigma^{m}\psi^{(1)}=[q(x^0)]^{2}\sigma^{m}_{21}
\end{equation}\\
where $\sigma^{m}$ (m=1,2,3) are Pauli Matrices and $\sigma^{0}=1$.\\

Now, we construct space time tetrad differently from  [13]  and others
in the following way.\\
\begin{equation}\left(
    \begin{array}{c}
      e^{m \nu}=e^{\nu m}  \\
    \end{array}
  \right)=\left(
            \begin{array}{c}
              e^{0m} \\
              e^{1m}  \\
              e^{2m}  \\
              e^{3m}  \\
            \end{array}
          \right)=\left(
                    \begin{array}{c}
                      \frac{1}{2}q^{2}(\sigma^{m}_{11}+\sigma^{m}_{22}) \\
                      \frac{1}{2}q^{2}(\sigma^{m}_{12}+\sigma^{m}_{21}) \\
                      \frac{-i}{2}q^{2}(\sigma^{m}_{12}-\sigma^{m}_{21}) \\
                      \frac{1}{2}q^{2}(\sigma^{m}_{11}-\sigma^{m}_{22}) \\
                    \end{array}
                  \right)=\left(
                            \begin{array}{c}
                              q^{2}\eta^{0m}  \\
                              q^{2}\eta^{1m}  \\
                              q^{2}\eta^{2m}  \\
                              q^{2}\eta^{3m} \\
                            \end{array}
                          \right) \end{equation}
where $\eta_{\mu \nu}$ is the Minkowski metric tensor with
signature $(+,-,-,-)$.\\

Thus \begin{equation}  e^{m \nu}=e^{\nu m}=q^2\eta^{m \nu}~
=q^2\eta^{  \nu m}~ and ~e_m^\mu =q^2 \eta_m =q^2 \delta_{\mu m}
\end{equation}

 Here, we get $e^\mu_m$ from $e^{m \nu}$ by
transforming the spinors according to (5).  We define a tensor
$G^{\mu \nu}$ by
\begin{equation}G^{\mu \nu}=e^\mu_m e^{m\nu }\end{equation}
 Therefore, we have \begin{equation}G^{\mu \nu}=q^4 \eta^{\mu \nu}\end{equation}
 Now setting \begin{equation}G^{\mu \nu}\equiv G=ggg \equiv(g^{\rho \sigma})
 (g^{\rho \sigma})(g^{\rho \sigma}),\end{equation}
 we get \begin{equation}(ggg)^{\mu \nu}=q^4 \eta^{\mu \nu}\end{equation}
 A solution of this equation is given by  \begin{equation}
 g^{\mu \nu}=q^{\frac{4}{3}}
 \eta^{\mu \nu},~consequently,~g_{\mu \nu}=q^{-\frac{4}{3}}
 \eta_{\mu \nu}\end{equation}
 which is taken as the metric tensor constructed from the two
independent spinors.  By using the relations (26) and (29) we get
\begin{equation}g_{\mu \nu}=g(x^0)\eta_{\mu \nu}\end{equation}
 This metric tensor gives rise to the same geometric structure (given
in (28)) of the manifold that emerged by the appearance of the
Weyl spinors having the sympletic structure in the abstract
spinor space. That is, the spinors of the flat space can produce
the curved spacetime which is the background FRW universe being
equivalent to the geometric structure (38).
 It will be a noteworthy fact that we can also arrive at a Finsler
spacetime structure of the manifold apart from the Riemannian
structure as above. In fact, the Finsler space that we shall get
has the above space time with metric tensor (38) as the associated
Riemannian spacetime of it. The fundamental function of the
Finsler space is defined here as  \begin{equation}F^4=G_{\mu \nu
\rho \sigma}(x) \nu^\mu \nu^\nu \nu^\rho \nu^\sigma\end{equation}
 or, \begin{equation}ds^4=G_{\mu \nu \rho \sigma}(x) dx^\mu dx^\nu
dx^\rho dx^\sigma \end{equation}
 The Finsler space introduced here is, in fact, in accord with
Riemann's original suggestion that the positive fourth root of a
fourth order differential form might serve as a metric function
(Riemann,1854). Here we take  \begin{equation}G_{\mu \nu \rho
\sigma}=G_{\mu \nu} G_{\rho \sigma}=q^{-8} \eta_{\mu \nu}
\eta_{\rho \sigma}\end{equation}  by using (34). Consequently, we
have from (40)
\begin{equation} ds^4=q^{-8} \eta_{\mu \nu} \eta_{\rho \sigma} dx^\mu
dx^\nu dx^\rho dx^\sigma=(q^{-4 } \eta_{\mu \nu} dx^\mu
dx^\nu)^2\end{equation}
 Now there are two possibilities from this relation :

(i)~\[ ds^2=q^{-4} \eta_{\mu \nu} dx^\mu dx^\nu,  \] i.e., the tensor
$G_{\mu \nu}$ is given by \[ G_{\mu \nu}=q^{-4} \eta_{\mu \nu}.\]
This leads to the Riemannian metric tensor (38).

(ii) On the other hand
, we have
 \begin{equation} ds^2=\theta(dx^2)q^{-4} \eta_{\mu \nu} dx^\mu
dx^\nu\end{equation}  where \begin{equation} dx^2=\eta_{\mu \nu}
dx^\mu dx^\nu\end{equation}  and
\begin{equation} \theta(z)=1,~z\geq0~;~
  \theta(z)
  =-1,z<0\end{equation}
 From (43), we get a tensor
\begin{equation}G_{\mu \nu}=\theta(dx^2)q^{-4} \eta_{\mu
\nu}\end{equation}  Writing as before, $(G_{\mu \nu})\equiv
G=ggg$ and noting that
\begin{equation} {\theta(dx^2)}^3=\theta(dx^2)=\theta(\nu^2)\end{equation}
 where\begin{equation} \nu^\mu \equiv
\frac{dx^\mu}{ds},\end{equation}  we have \begin{equation} g_{\mu
\nu}=\theta(\nu^2)g(x^0)\eta_{\mu \nu}\end{equation}  and
consequently, \begin{equation} F^2=g_{\mu \nu}\nu^\mu
\nu^\nu=\theta(\nu^2)g(x^0)\eta_{\mu \nu} \nu^\mu
\nu^\nu=g(x^0)\theta(\nu^2) \nu^2.\end{equation}  This is a
Finsler space which was considered by [17]  ( see also [18] )  as the anisotropic spacetime of hadronic matter
extension. Here,this spacetime has been constructed from the
tetrad fields by considering their correspondence with the
two-dimensional spiner space endowed with a symplectic structure.
It is to be noted that the Riemannian space time with the metric
tensor (38) is the associated Riemannian manifold of the Finsler
space. This Riemannian spacetime structure is the emergent
gravity and it is, in fact, the cosmological background spacetime
of the universe,as shown earlier. Once the structureless spacetime
manifold transformed spontaneously into Finslerian manifold (with
an associated Riemannian manifold), the spinor fields must
consequently satisfy a field equation in that space. This equation
has been proposed by [17]  for the four-dimensional
bispinors. The equation was derived from a classical field
equation by the quantization of  the space and time at the Planck-
order scale. This quantization of space and time is also a
consequence of the assumption of the minimum length and time.
Also, the minimum length and time follow from the generalized
uncertainty principle. We now discuss this GUP in next section.
\\

\section{ {\LARGE GENERALIZED UNCERTAINTY OPRINCIPLE:}}

Modifications of the Heisenberg uncertainty principle near the
planck scale are the GUP which are motivated by string
theory,black hole physics,doubly special relativity and other
theories of quantum gravity.Physical implications such as
discreteness of space time have been considered by several
authors [see [19] , [20] and references there
in]. Here we examine the implication of GUP as proposed by [21]   because in there the origin of the additional
term making the modification has been discussed, and thus this
can be regarded as independent of any model.They proposed the GUP
as\begin{equation} \delta x=\delta x_{H}+\delta
x_{G}\end{equation} where \begin{equation}\delta x_{H} \geq
\frac{\hbar}{2\delta p}\end{equation} and \begin{equation}\delta
x_{G}\geq \frac{2L^{2}_{p}\delta p}{\hbar}\end{equation} where
$\delta p$ is the momentum uncertainty of the electron and
$L_{p}$ is the planck length.$\delta x_{H}$ is the Heisenberg
uncertainty of position of electron and $\delta x_{G}$ is the
additional term to the uncertainty of the position. Derivation of
this additional term $\delta x_{G}$ has been made by them in four
heuristic ways, two of which are based on Newtonian theory of
gravitation where as the other two are on consideration of
general relativity theory.

In fact, the position uncertainty of electron $\Delta x_{G}$ was
shown to arise out
 from the gravitation interaction between photon as quanta and electron.It is to be
 noted that the electron momentum uncertainty $\Delta p$ must be of the order of
 interacting photon momentum $p=\frac{E}{c}=\frac{\hbar}{\lambda}$ where $E$ is the
 energy of photon and $\lambda$ is the photon wave length. Thus\begin{equation}
 \delta p\simeq p=\frac{\hbar}{\lambda}\end{equation}

 Consequently, we have from (52),\begin{equation} \delta x_{H}\geq \frac{\lambda}{2}\end{equation}
 This was Heisenberg's contention that one can not obtain better precision in the
 position of electron than $\lambda$ when measured with an electromagnetic wave
  of that wave length.
 That is the shorter wave length being responsible of good resolution cab make
 the position measurement of electron more precise.
 On the other hand, we have from (53) and (54),
\begin{equation}\delta {x_{G}}\geq \frac{L^{2}_{p}}{\lambda}\end{equation}
 \\
 Thus we see that with the shorter wave length $\lambda$ which corresponds to
 the higher energy and momentum of photon,
 the gravitational photon field interacting with electron makes its position
  less precise. Physically, such situation
 one should expect and can think that the additional uncertainty term $\Delta x_{G}$
  is indeed, due to gravity and consequently
 related to the curved space time structure.\\
 Gao (2010), in a reverse application of GUP (51) of Adler and Santiago, has argued
 the energy and momentum of a particle can
 change the geometric structure of space time. There it is shown that the energy $E\simeq pc$
 in a region with size $L$ can
 change the proper size of the region because of the positional uncertainty $\Delta x_{G}$
 as in (53). The change is by an amount,
 \begin{equation}\delta L\simeq \frac{L_{p}T_{p}E}{\hbar}\end{equation}
 \\
 where $T_{p}$ is the planck time.This implies that a flat space time transformed to the
 curved one by the energy momentum in the region.
 \\
 Consequence of the GUP (51) is that the position uncertainty $\Delta x$ has a minimum
  \begin{equation}
 \delta x_{min}\simeq 2L_{p}\end{equation}

 Also the similar uncertainty principle (for time uncertainty)
\begin{equation} \delta t \geq \frac{\hbar}{2 \Delta E} + \frac{2 T^{2}_{p}
\Delta E}{\hbar}\end{equation}
 leads to a minimum  time uncertainty,

\begin{equation}
 \delta t_{min}\simeq 2T_{p}\end{equation}
 The implication of the minimum size of space time is that the interval shorter than
 planck scale is physically meaningless,
 that is the measurements can not be possible in interval shorter than this scale.
 In other words space time is quantized at the planck scale level
 [20].
 \\
 Now in the present case,the structureless space time manifold is transformed spontaneously
 into a curved one, the cosmological background space time
 of the universe.Thus gravity emerged and due to GUP the space time became quantized with
 the minimum planck order length and time. On the contrary, the
minimum length and time or quantized space time which is the
consequence of QM can give rise to gravity due to emergent curved
spacetime manifold.  So it can be thought as a phenomenon of
emergence of both QM and gravity. Here we see that it is due to
the appearance of Weyl spiners dependent only on time in the flat
spacetime manifold,and having a sympletic structure in the
abstract spiner space.

\section{\LARGE
FIELD EQUATION AND MASSIVE PARTICLES:\\}

As in [17]  the field equation for the two-dimensional
'bispinor' in the Finsler space can be derived for the massless
case.(as also for massive bispinor fields). This equation for the
bispinor  $\Psi \equiv (\Psi_{\alpha \beta})$ is (in the natural
unit $\hbar=c=1$)  \begin{equation}i \sigma_{\alpha
\beta^\prime}^\mu(x)
\partial_\mu \Psi_{\beta^\prime \beta}(x,\nu)\mp i\sigma_{\beta
\beta^\prime}^\mu(x) \gamma_{h \mu}^l(x,\nu) \nu^h \partial_l^\prime
\Psi_{\alpha \beta^\prime}(x,\nu)=0,\end{equation}
 where $\gamma_{h \mu}^l(x,\nu)$ are Christoffel symbol of second kind
,and  \begin{equation}\partial_{\mu} \equiv \frac{\partial}{\partial
x^\mu},~~~\partial_l^{\prime} \equiv \frac{\partial}{\partial
\nu^l}.\end{equation}
 Pauli matrices $\sigma^i (x)$(i=1,2,3) and $\sigma^0 (x)$ are
related to the flat space Pauli matrices $\sigma^i$ and
$\sigma^0=1$ by the vierbein   for the associated Riemannian
manifold of the present Finsler space. The above equation (61)
can be written as
\[ i \sigma^\mu_{\alpha \beta^\prime}(x)\{\partial_\mu
\Psi_{\beta^\prime \beta}\mp \sigma_{\beta^\prime \nu}^{\mu^{-1}}(x)
\sigma^\mu_{\beta \delta} (x) \gamma_{h \mu}^l (x,\nu)\nu^h
\partial_l^\prime \Psi_{\nu \delta}\}=0,\]
  or,\begin{equation}i \sigma_{\alpha \beta^\prime}^\mu (x) {\partial_\mu \Psi_
  {\beta^\prime \beta}(x,\nu)\mp
   A^l_{\mu \beta^\prime \beta \nu \delta}(x,\nu) \partial^{\prime}_l \Psi_
   {\nu \delta}(x,\nu)}=0,\end{equation}
 where \begin{equation}A^l_{\mu \beta^\prime \beta \nu
\delta}(x,\nu)=\sigma_{\beta^\prime \nu}^{{\mu}^{-1}}(x)
\sigma^\mu_{\beta \delta}(x) \gamma^l_{h \mu}(x,\nu)
\nu^h . \end{equation}
 That is, the equation can be written in compact form as
\begin{equation}i \sigma^\mu \bigtriangledown_\mu
\Psi(x,\nu)=0,\end{equation}
 where \begin{equation}\bigtriangledown_\mu=\partial_\mu \mp A^l_\mu (x,\nu)
\partial_l^\prime,\end{equation}
 is the generalized derivative.
 Now, for the Finsler space (50), $g_{\mu \nu}$ as in (49) is not
Finslerian metric. On the other hand, the Finsler metric is given
by \[ \widehat{g}_{ij}=\frac{\partial^2 F^2}{2 \partial \nu^i
\partial \nu^j}.\]  The Christoffel symbols $\gamma_{h \nu}^l(x,\nu)$
are derived from the metric tensor $\widehat{g}_{ij}$ but it can
be proved that the relation
\begin{equation}\gamma^i_{jk} \nu^j \nu^k=\widehat{\gamma}_{jk}^i \nu^j
\nu^k , \end{equation}
 holds good for the present Finsler space, where the Christoffel
symbols $\widehat{\gamma}_{jk}^i$ are derived from $g_{ij}$.
Because of this relation, we can use $\gamma_{jk}^i$ in the above
field equation in stead of $\gamma_{jk}^i$. We can obtain
$\gamma_{h\mu}^l$ as

\begin{equation}\widehat{\gamma}_{h \mu}^l=\frac{1}{2}\left[\delta_h^l \frac{\partial \ln
g(x^0)}{\partial x^\mu}+\delta_\mu^l \frac{\partial \ln
g(x^0)}{\partial x^h}-\eta^{l \beta}\eta_{h \mu} \frac{\partial
\ln g(x^0)}{\partial x^\beta}\right].\end{equation}  The equation
(63) retain its form with the flat space Pauli matrices
$\sigma^i$(i=1,2,3) and $\sigma^0$ because of the vierbein for
the spacetime  (28), and therefore, using (68) we have  \[ i
\sigma^\mu A_\mu^l (x,\nu)
\partial_l^\prime \Psi \equiv i \sigma_{\alpha \beta^\prime}^\mu
A^l_{\mu \beta^\prime \beta \nu \delta} \partial_l^\prime
\Psi_{\nu \delta}=\frac{i}{2}\sigma^\mu_{\alpha
\beta^\prime}\sigma^\mu_{\beta^\prime \nu}\sigma^{\mu T}_{\beta
\delta}[\nu^l \delta_\mu^0+\delta_\mu^l \nu^0-\eta^{l_0}\eta_{h
\mu}\nu^h]\frac{dl_n g(x^0)}{dx^0}\sigma_l^\prime \Psi_{\nu
\delta}\] \begin{equation}=\frac{i}{2}\delta_{\alpha \nu}
\sigma^{\mu T}_{\beta \delta}
\frac{g^\prime(x^0)}{g(x^0)}[\delta_\mu^l \nu^0-\eta^{l0} \eta_{h
\mu}\nu^h]\partial^\prime_l \Phi_\delta \Psi_\nu,
\end{equation}
 where the bispiner $\Psi_{\nu \delta}$ has been decomposed as
\begin{equation} \Psi_{\nu \delta }(x,\nu)=\Phi_\delta(\nu)
\Psi_\nu(x), \end{equation}
 $\Phi_\delta(\nu)$ being a two dimensional spinor which is supposed
to be a homogeneous function of\\ $\nu=(\nu^0,\nu^1,\nu^2,\nu^3)$
of degree zero,  i.e.,\begin{equation}\Sigma^{3}_{l=1} \nu^l
\partial_l^\prime \Phi_\delta(\nu)=0 . \end{equation}
 Then we have
  \begin{equation}\mp i\sigma_{\alpha \beta^\prime}^\mu A^l_{\mu \beta^\prime \beta \nu
  \delta}\partial^\prime_l
  \Psi_{\nu \delta}=\mp \frac{i}{2}\Psi_\alpha \frac{dl_n g(x^0)}{dx^0}\Sigma^{3}_{l=1}
   \sigma_{\beta \delta^{iT}}
  (\nu^0 \partial_i+\nu^i \partial_0)\phi_\delta= \frac{3ig^\prime(x^0)}{4g(x^0)}\Psi_\alpha
  \Phi_\beta , \end{equation}
 if \begin{equation}\Sigma^{3}_{l=1} {\sigma_{\beta \delta}^{iT}(\nu^0 \partial_i+\nu^i
\partial_0)\Phi_\delta(\nu)}=\mp
\frac{3}{2}\phi_\beta(\nu) . \end{equation}
 Therefore, from (65),~(66) and (72) we have
\[ \sigma^\mu_{\alpha \beta^\prime}\partial_\mu
\Psi_{\beta^{\prime}}(x)\Phi_\beta(\nu)+\frac{3g^\prime(x^0)}{4g(x^0)}\Psi_\alpha(x)
\Phi_\beta(\nu)=0,\]
 or, \begin{equation} \sigma_{\alpha \beta^\prime}^\mu \partial_\mu
\Psi_{\beta^\prime}(x)+\frac{3g^\prime(x^0)}{g(x^0)}\Psi_\alpha(x)=0\end{equation}
 That is, \[ ~~~~~~~~~~~~~~~~~~~~~\sigma^\mu \partial_\mu
\Psi(x)-\frac{3g^\prime(x^0)}{4g(x^0)}\Psi(x)=0.
~~~~~~~~~~~~~~~~~~~~~~~~~~~~~~~~~~~~~~~~~~~~~~~~~~~~~~~~~~~~~~~~~~~~~~~(74a)\]
This equation is the same as (21) for the time dependent spinors.
Also, we can have Dirac equation for massive spiner field in the
associated Riemannian spacetime manifold of the Finsler space as
in De (2003). The field equation for the bispiner
$\Psi(x,\nu)={\Psi_{\alpha \beta}(x,\nu)}$ in the Finsler space is
\begin{equation} i\hbar{\gamma^\mu_{\alpha \beta^\prime}(x)\partial_\mu \Psi
_{\beta^\prime \beta}(x,\nu)-\gamma^\mu_{\beta
\beta^\prime}(x)\gamma^l_{h\mu}(x,\nu)\nu^h \partial^\prime_l
\Psi_{\alpha \beta^\prime}(x,\nu)}=mc \Psi_{\alpha
\beta}(x,\nu),\end{equation}
 where $\gamma^\mu_{\alpha \beta^\prime}(x)$ are related to the flat space Dirac matrices
$\gamma^\mu$ by the vierbein for the spacetime metric tensor (28).

With the decomposition  \begin{equation} \Psi(x,\nu)=\Psi_1(x)
\Phi^T(\nu)+\Psi_2(x)\Phi{^c}^T(\nu),\end{equation} where $\Psi_1$
and $\Psi_2$ are eigenstates of $\nu^0$ with eigenvalues +1 and -1
respectively, we have the following Dirac equation for
$\Psi(x,\nu)$:
\begin{equation} i\hbar\left(\gamma^\mu
\partial_\mu-\frac{3b_0}{2}\xi(t)\nu^0\right)\Psi(x,\nu)=\frac{c}{e(t)}(m+M
\xi(t) e(t))\Psi(x,\nu), \end{equation}
 if $\Phi(\nu)$ and $\Phi^c(\nu)$ satisfy the following equations
\begin{equation}i\hbar \gamma^\mu \gamma^l_{h\mu}\gamma^h \partial_l^\prime
\Phi(\nu)=\left(Mc-\frac{3i\hbar b_0}{2}\right)\Phi(\nu),
\end{equation}
\[~~~~~~~~~~~~~~~~~~~~~~~~~i\hbar \gamma^\mu \gamma^l_{h\mu}\nu^h \partial_l^\prime
\Phi^c(\nu)=\left(Mc+\frac{3i\hbar b_0}{2}\right)\Phi^c(\nu).
~~~~~~~~~~~~~~~~~~~~~~~~~~~~~~~~~~~~~~~~~~~~~~~~~(78a)\]
 If $\Phi(\nu)$ and $\Phi^c(\nu)$ are homogeneous function of
$\nu=(\nu^0,\nu^1,\nu^2,\nu^3)$ of degree zero ,then the equations
(78) and (78a) become \begin{equation} i\hbar b_0 \Sigma^{3}_{l=1}
\nu^l(\nu^l \frac{\partial}{\partial \nu^0}+\nu^0
\frac{\partial}{\partial \nu^l})\Phi(\nu)=-\left(Mc-\frac{3i\hbar
b_0}{2}\right)\Phi(\nu), \end{equation}

\[~~~~~~~~i\hbar b_0 \Sigma^{3}_{l=1} \nu^l(\nu^l \frac{\partial}{\partial \nu^0}+\nu^0
\frac{\partial}{\partial
\nu^l})\Phi^c(\nu)=-\left(Mc+\frac{3i\hbar
b_0}{2}\right)\Phi^c(\nu).
~~~~~~~~~~~~~~~~~~~~~~~~~~~~~~~~~~~~~~~~~~~~~(79a)\]
 Here, \begin{equation} e(t)=\frac{1}{\sqrt{g(x^0)}},~~~~~2b_0
\xi(t)=\frac{g^\prime(x^0)}{g{(x^0)}}, \end{equation}
 and m is the inherent mass of the spinor.  Also, here M appears as a
constant in the process of separation of the field equation (75).
$\Psi(x,\nu)$ satisfies the Dirac equation (77) in the associated
Riemannian space which is, in fact, a space time conformal to
Minkowski flat space time.The Dirac equation for FRW spacetime
can be obtained by the pure time transformation (29). If $M\neq0$,
the additional mass term is time dependent.
 The mass of the spinor is given by
\begin{equation} m+M\xi(t)e(t)=m+\frac{M}{b_0 c}H(T),\end{equation}
 where \[ H(T)=\frac{R^\prime(T)}{R(T)}, \] is the Hubble's function.For
the case M=0, it can be seen that the $\nu$ dependent spiners
$\Phi(\nu)$ and $\Phi^c(\nu)$ in the decomposition of
$\Psi(x,\nu)$ can give rise to an additional quantum number if it
is the field of the constituent-particle in the hadron
configuration.This quantum number can generate the interval
symmetry of hadrons [18, 22 ]. The parameter
$b_0$ of the spacetime in the epoch dependent mass term in (81) is
found to be connected with M. This relation has been obtained in
the derivation of solutions for $\phi(\nu)$ and $\phi^{c}(\nu)$
of the equations of 79 and 79(a) [17]. It is shown there
that $M$ must be proportional to $b_{0}$ as follows:\\
\begin{equation}(0.832)\hbar b_{0}=Mc . \end{equation}
Then the mass term in (81) becomes\\
\begin{equation}m+(0.832)\frac{\hbar}{c^2} H(T). \end{equation}
Then the parameter $b_{0}$ is in fact connected to the expansion
of the universe.\\ For example $g(x^0)=(b_{0}x_{0})^{4}$. The
scale factor and Hubble's function
respectively found to be\\
\begin{equation}R(T)=(3b_{0}cT)^{\frac{2}{3}}, \end{equation}
and
\begin{equation}H(T)=\frac{2}{3T}.  \end{equation}
Thus we have an epoch dependent mass term apart from the inherent
mass of the spinor even for massless particles like neutrinos.
This epoch dependent mass which is of the order of planck mass at
the Planck order time (the minimum time as obtained from GUP)
becomes negligible at the present epoch of the universe. These
highly massive particles give rise to the gravitation by
fulfilling the Einstein equivalence with their mass distribution
in the emerged curved space time. The inherent masses of the
particles might be achievable from self intersection phenomenon
or otherwise [23]. Actually these highly massive
particles can deliver necessary mass density in the very early
universe [24].

\section{THE SCENARIO AND CONCLUDING REMARKS }

We have seen that due to the emerged gravity the Heisenberg
uncertainty principle is to be modified into GUP, which in turn
lead to the discreteness of space time with the minimum time and
length uncertainty, and vice versa. As an implication of the
discreteness of the space time or consequent GUP, Gao [25] also
has shown that flat space becomes curved by the energy momentum
contained in it. This provides a deeper meaning for the Einstein
Equivalence principle in G R. On the basis of this principle it
is there pointed out that Einstein field equation can be derived
by imposing some reasonable conditions on the Riemannian
symmetric second rank tensor constructed from the metric of the
space time. and its derivative and on energy momentum (i.e, the
conservation of energy and momentum) i.e., the emerged curved
manifold together with the highly massive particles there in all
emerged instantaneously and simultaneously with the Quantum Weyl
particles dependent only on time and
 with the symplectic conditions satisfied by them in the abstract spinor space can be regarded
 as the origin of the expanding universe which is in fact, the associated Riemannian manifold
 of the emerged Finsler space. In a sense, both Q M and gravity as derived by G R are
  fundamental
 as well as emergent. The phenomenon of emergence rather, the unified emergence occurs
  instantaneously.
 The instantaneousness is in fact, the minimum time (as obtained from GUP) which is of
 the order of
 Planck time and the fact is that it is indivisible. Thus the minimum time is only conceivable
 as zero time or instantaneous. After that Planck order time, the time may be regarded as
  continuous
 because we need not have to consider any phenomenon in later time in the minimum
 inconceivable and
 indivisible discrete time.\\
 In the scenario described above we see a parameter $b_{0}$ which is
related with the
 space time structure. This parameter specifies the time dependence of Weyl spinors as well as
 consequent structure of the emergent space time manifold. Obviously it appears as the
  scale factor
 of the expanding universe. Therefore it determines the age of the universe from the
 scale factor
 of FRW space time. Again we see that $b_{0}$ is linearly related to the mass term parameter $M$.
 Therefore$ M $ is not a new parameter. On the other hand, if $M$ is taken as to be
  Planck order mass,
 then $b_{0}$ is fixed by it, and the age of the universe is specified.Thus $b_{0}$
  and $M$ are no new
 parameters, they are derivable from the fundamental constants $\hbar$,$c$ and the
 gravitational constant
 $G$.\\
 Finally we like to say that the gravity may be quantized in the present framework
 because the metric
  structure of the space time is here constructed from the spinors which can be
  subjected to quantization
  as in Q M.
  \\
  \\

\section*{Acknowledgments} FR is thankful to the
Inter-University Centre for Astronomy and Astrophysics (IUCAA),
India for providing Visiting Associateship. We are thankful to the referee for his comments. FR is also grateful to the DST, Govt. of India, for financial support under
PURSE programme.

 \end{document}